\documentstyle[12pt]{article}
\begin{document}
\title{RELATIVISTIC FLUIDS AND THE PHYSICS OF GRAVITATIONAL COLLAPSE}
\author{L. Herrera$^1$\thanks{On leave from UCV, Caracas, Venezuela, e-mail: lherrera@usal.es}\\
{$^1$Departamento de  de F\'{\i}sica Teorica e Historia de la  F\'{\i}sica,}\\
Universidad del Pais Vasco, Bilbao, Spain}
\maketitle
\begin{abstract}
 We present  a discussion on some physical aspects of gravitational collapse which is  based on  a list of questions  related to relevant issues in the study of that phenomenon.  Providing answers to those questions we  bring out the role played by different physical processes in the dynamics of spherical collapse.
\end{abstract}
\date{}

\section{INTRODUCTION}
The gravitational collapse of massive stars and its resulting
remnant (neutron star or black hole) represent one
of the few observable scenarios where general relativity is
expected to play a relevant role. Therefore, a detailed
description of gravitational collapse and the modelling of
the structure of compact objects under a variety of conditions
remain among the most interesting problems that
general relativity has to deal with. This fact explains the
great attraction that these problems exert on the community
of the relativists. 

Motivated by the above arguments, we present here a discussion on self-gravitating relativistic fluids, emphasizing  the role played by some specific physical phenomena like local anisotropy of pressure, dissipation and electric charge. The discussion will be centred on the answers to a list of questions  related to essential aspects  of the dynamics of gravitational collapse.

These questions  are:

\begin{itemize}
\item Why is the gravitational
collapse a highly dissipative process? 
\item  Is there  life between quasi--equilibrium and non--equilibrium?
\item What happens if we relax  the Pascal principle? (and why?)
\item How do  electric charge and dissipation affect the evolution of massive stars?
\end{itemize}

\section{WHY IS THE GRAVITATIONAL COLLAPSE A HIGHLY DISSIPATIVE PROCESS?}
In order to answer to this question let us present some very rough estimates. The important point is that no matter how much we  refine them, the answer will be essentially the same.

So, let us consider a massive star which collapses producing a one solar mass neutron star, and let us further assume that this process proceeds adiabatically. 

Now  a rough calculation for the gravitational binding energy of a one solar mass of $10$ Km radius, given by 
\begin{equation}
E\approx-\frac{G M^2}{R}\approx -\frac{6.6 \times 10^{-8}g^{-1}cm^3 s^{-2}10^{66}g^2}{10^6 cm}
\end{equation}
with
$M=M_{\odot}\approx 2. 10^{33} g$;\quad \ $R\approx 10 Km$

produces
\begin{equation}
E\approx -10^{53} erg.
\end{equation}

Then  at least $10^{53} erg$ has to be generated  during the collapse, therefore if we assume that collapse  proceeds adiabatically then  internal energy ($E_{in}$) has to increase by that amount.

Next, assuming 
\begin{equation}
E_{in}\approx kT,
\end{equation}

where the Boltzman constant is  $k\approx 1.3 \times 10^{-16} erg. K^{-1}$.
Then the temperature of such a system should rise to the absurd value
\begin{equation}
T\approx 10^{69} K,
\end{equation}
 implying that after all it should radiate with a luminosity 

\begin{equation}
L\approx 10^{283}erg.s^{-1},
\end{equation}
which would evaporate our object at a time scale of the order of

\begin{equation}
t\approx (10)^{-230}s.
\end{equation}

This sequence of ``absurdities''  is  produced by the original assumption of adiabaticity. The fact is that  in the process of collapse the system has to get rid of a huge amount of energy ($\approx 10^{53} erg$ ). Accordingly dissipation and radiative transport within the star are bound to play an essential role in the dynamics of collapse.

We shall come back to this point in the last section.

\section{IS THERE  LIFE BETWEEN \\QUASI--EQUILIBRIUM AND \\NON--EQUILIBRIUM?}
 In the study of gravitational collapse it is usual to appeal to  numerical methods, which   allow for
considering more realistic equations of state. However,  the obtained results,
in general, are restricted and highly model dependent. Also, specific difficulties, associated with numerical solutions of partial
differential equations in presence of shocks, further complicate  the
problem.

On the other hand, analytical solutions although more suitable for a
general discussion, are 
found, either for too simplistic equations of state and/or under additional
heuristic assumptions whose justification is usually uncertain.

On the grounds of the  comments above, it is justified to introduce  seminumerical techniques, which  may be regarded
as a ``compromise'' between the analytical and numerical approaches. These techniques developed in \cite{HJR}, \cite{FCP} and \cite{5bis} are based on a general algorithm for modelling self--gravitating spheres out of equilibrium.

The proposed method, starting from any interior (analytical) static spherically
symmetric (``seed'') solution to Einstein equations, leads to a system of ordinary differential equations for quantities evaluated at the boundary surface of the fluid distribution,
whose solution (numerical), allows for modelling, dynamic, self-gravitating spheres, whose static limit is the original ``seed'' solution.

The approach is based on the introduction of a set of conveniently defined ``effective'' variables (effective pressure and energy density) and an heuristic ansatz on the latter, whose rationale and justification become intelligible within the context of the post-quasistatic approximation. In the quasistatic approximation, the effective variables coincide with the corresponding physical variables (pressure and density) and  therefore the method may be regarded as an iterative method
with each consecutive step corresponding to a stronger departure from equilibrium.

Such a seminumerical technique requires the possibility to approach the non--equilibrium state by means of successive approximations, implying an affirmative answer to the question posed  in the title of this section.

In order to show in some detail the basic features of the proposed approach, let us first give a very brief description of the Bondi  formalism for the description of spherical collapse in non--comoving coordinates (see \cite{B} for details).

The line element is given in Schwarzschild--like coordinates by
\begin{equation}
ds^2=e^{\nu} dt^2 - e^{\lambda} dr^2 -
r^2 \left( d\theta^2 + sin^2\theta d\phi^2 \right),
\label{metric}
\end{equation}
where $\nu(t,r)$ and $\lambda(t,r)$ are functions of their arguments.

The metric given in (\ref{metric}) should satisfy  the Einstein equations

\begin{equation}
G^\nu_\mu=8\pi T^\nu_\mu.
\label{Efeq}
\end{equation}
\noindent
In order to give physical significance to the $T^{\mu}_{\nu}$ components
we apply the Bondi approach \cite{B}.

\noindent
Thus, following Bondi, let us introduce purely locally Minkowski
coordinates ($\tau, x, y, z$)

$$d\tau=e^{\nu/2}dt\,;\qquad\,dx=e^{\lambda/2}dr\,;\qquad\,
dy=rd\theta\,;\qquad\, dz=rsin\theta d\phi.$$

\noindent
Then, denoting the Minkowski components of the energy tensor by a bar,
we have

$$\bar T^0_0=T^0_0\,;\qquad\,
\bar T^1_1=T^1_1\,;\qquad\,\bar T^2_2=T^2_2\,;\qquad\,
\bar T^3_3=T^3_3\,;\qquad\,\bar T_{01}=e^{-(\nu+\lambda)/2}T_{01}.$$

\noindent
Next, we suppose that when viewed by an observer moving relative to these
coordinates with proper velocity $\omega$ in the radial direction, the physical
content  of space consists of an anisotropic fluid of energy density $\rho$,
radial pressure $P_r$, tangential pressure $P_\bot$,  radial heat flux
$ q$ (which models dissipation in the diffusion approximation) and unpolarized radiation of energy density $\epsilon$
traveling in the radial direction (which models dissipation in the free streaming regime). Thus, when viewed by this (comoving with the fluid) 
observer the covariant tensor in
Minkowski coordinates is

\[ \left(\begin{array}{cccc} 
\rho + \epsilon    &  - q -\epsilon  &   0     &   0    \\
- q - \epsilon &  P_r + \epsilon    &   0     &   0    \\
0       &   0       & P_\bot  &   0    \\
0       &   0       &   0     &   P_\bot
\end{array} \right). \]
Then a Lorentz transformation gives us the components of $T^\mu_\nu$ in terms of the physical variables defined by the comoving Minkowskian observer.

Thus Einstein equations read:

\begin{equation}
\frac{\rho + P_r \omega^2 }{1 - \omega^2} +\frac{2\omega
q}{1-\omega^2}+\frac{\epsilon(1+\omega)}{1-\omega}=-\frac{1}{8 \pi}\Biggl\{-\frac{1}{r^2}+e^{-\lambda}
\left(\frac{1}{r^2}-\frac{\lambda'}{r} \right)\Biggr\},
\label{fieq00}
\end{equation}

\begin{equation}
\frac{ P_r + \rho \omega^2}{1 - \omega^2} +
\frac{2\omega
q}{1-\omega^2}+\frac{\epsilon(1+\omega)}{1-\omega}=-\frac{1}{8 \pi}\Biggl\{\frac{1}{r^2} - e^{-\lambda}
\left(\frac{1}{r^2}+\frac{\nu'}{r}\right)\Biggr\},
\label{fieq11}
\end{equation}

\begin{eqnarray}
P_\bot = -\frac{1}{8 \pi}\Biggl\{\frac{e^{-\nu}}{4}\left(2\ddot\lambda+
\dot\lambda(\dot\lambda-\dot\nu)\right) \nonumber \\
 - \frac{e^{-\lambda}}{4}
\left(2\nu''+\nu'^2 -
\lambda'\nu' + 2\frac{\nu' - \lambda'}{r}\right)\Biggr\},
\label{fieq2233}
\end{eqnarray}

\begin{equation}
\frac{(\rho + P_r) \omega e^{\frac{\lambda +\nu}{2}}}{1 - \omega^2} +\frac{q e^{\frac{\lambda+\nu}{2}}}{1-\omega ^2}(1+\omega
^2)+\frac{e^{\frac{\lambda +\nu}{2}}
\epsilon(1+\omega)}{1-\omega}=-\frac{\dot\lambda}{8 \pi r},
\label{fieq01}
\end{equation}
where dot  denotes diferentiation with respect to $t$ and prime denotes $r$ derivatives.

This is a system of four equations in partial derivatives for  eight unknown functions: $\nu, \lambda, \rho, P_r, P_\bot, \epsilon, q, \omega$. Accordingly additional information has to be given in order to integrate the system. Observe that  if the metric functions are completely determined,  the system (\ref{fieq00}--\ref{fieq01}) becomes an algebraic  system of four equations for six unknown functions, implying that  the system is not completely determined unless, infomation is provided through two additional equations. In  other words the complete knowledge of spacetime is not enough to reproduce all  physical properties of  the sources, a fact which is known but usually overlooked.

We shall assume our fluid distribution to be bounded by a spherical surface  $\Sigma$ outside of which we have the Vaydia spacetime. In order to avoid the existence of shells on $\Sigma$, we shall demand the first and second fundamental form to be continuous across it (Darmois).

It is sometimes useful to consider the ``conservation'' laws derived from the Bianchi identities, 
\begin{equation}
T^\mu_{\nu;\mu}=0.
\label{dTmn}
\end{equation}
which, even though  contain no further information than that found in Einstein equations may  provide  an interesting hindsight on the problem.

Thus for example in the static case the radial component of (\ref{dTmn}) reads

\begin{equation}
P'_r=-\frac{m + 4 \pi P_r r^3}{r \left(r - 2m\right)}\left(\rho+P_r\right)+
\frac{2\left(P_\bot-P_r\right)}{r},
\label{Prp}
\end{equation}
where $m$ is the mass function defined by 
\begin{equation}
m(r,t)=\frac{r}{2}(1-e^{-\lambda})= 4\pi \int_0^r{T_0^0 r^2 dr}.
\label{mT00n}
\end{equation}

Equation (\ref{Prp}) is the well known Tolman--Oppenheimer--Volkov equation (generalized for an anisotropic fluid)  whose physical meaning has been extensively discussed in literature.

Now, we may distinguish three possible regimes of evolution, namely: 
\begin{itemize}
\item Equilibrium. There is no evolution ($\omega=0$), metric and physical variables are independent on $t$ and equation (\ref{Prp}) is satisfied.
\item Quasi--equilibrium. The system evolves, but ``very slowly'', meaning that it changes on time scales much larger than hydrostatic time scale implying $\omega^2 \approx 0$. Thus equation (\ref{Prp}) is satisfied at all times and the evolution may be regarded as a sequence of equilibrium states. 
\item Non--equilibrium. The system changes on time scales of the order of (or smaller than) hydrostatic time scale.
\end{itemize}

Now the original question in the title of this section is equivalent to: Can we  approach the non--equilibrium by means of successive aproximations?  As we shall see, whenever the answer to this question is  affirmative,  we can describe the evolution of the system by means of a method implying resolution of ordinary differential equations instead of partial differential equations.
\subsection{Effective variables and the post--quasi--static approximation: The life between quasi--equilibrium and non--equilibrium}
Let us introduce the following  variables:

\begin{equation}
\tilde\rho=T^0_0= \frac{\rho + P_r \omega^2 }{1 - \omega^2} +
\frac{2 q\omega}{(1 - \omega^2)} + \epsilon\frac{(1+\omega)}{1-\omega},
\label{rhoeffec}
\end{equation}

\begin{equation}
\tilde P=-T^1_1=\frac{ P_r + \rho \omega^2}{1 - \omega^2} +
\frac{2 q \omega}{(1 - \omega^2)}+\epsilon\frac{(1+\omega)}{1-\omega}.
\label{peffec}
\end{equation}
which  we shall call respectively ``effective density'' and ``effective pressure''. Observe that in equilibrium and quasi--equilibrium the ``effective'' variables coincide with  the corresponding physical variables (energy density and radial pressure).

 Next it follows at once from the field equations that:

\begin{equation}
m =\frac{r}{2}(1-e^{-\lambda})= 4 \pi \int^{r}_{0}{r^2 \tilde\rho dr}, \\
\label{m1}
\end{equation}

\begin{equation}
\nu = \nu_\Sigma + \int^{r}_{r_\Sigma}\frac{2 (4 \pi r^3  \tilde
P+m)}{r(r-2m)} dr.
\label{nu}
\end{equation}
where $r=r_\Sigma$ is the equation of the boundary of the fluid distribution  and  subindex $\Sigma$ indicates that the quantity is evaluated at the boundary surface.

Thus it follows from the above that the radial dependence of metric functions is determined by the effective variables, this in turn implies that the radial dependence of a  given fluid distribution is the same in the equilibrium and the quasi--equilibrium regime.

Now if we assume that we can approach the non--equilibrium by successive approximations,  the question is: what could  we assume next for a regime  after  the quasi--equilibrium  (equation (\ref{Prp}) is not satisfied)? From the considerations above it appears that the only sensible assumption for such a regime would be that  metric variables share the same radial dependence that they have in equilibrium (or quasi--equilibrium). This assumption which we  shall call post--quasistatic--approximation allows us to develop an algorithm to describe (within such approximation) the evolution of self--gravitating spheres.

\subsubsection{The algorithm for modelling spheres out of equilibrium}

\begin{enumerate}
\item  Take an interior (seed) solution to Einstein equations, representing a fluid
distribution of matter in equilibrium, with a given

$$\rho_{st}=\rho(r)\,\qquad\, P_{r\, st}= P_{r}(r)$$

\item  Assume that the $r$--dependence of $\tilde P$ and $\tilde\rho$ is the
same as that of $P_{r\, st}$ and $\rho_{st}$, respectively, which implies that metric functions share the same radial dependence as that of the seed solution (the post--quasi--static--approximation).

\item  Using equations (\ref{m1}) and (\ref{nu}), with the $r$ dependence of
$\tilde P$ and $\tilde\rho$, one gets $m$ and $\nu$ up to some functions of
$t$, which have to be determined.

\item  For these functions of $t$ one has three ordinary differential equations
(hereafter referred to as surface equations), namely:
\begin{enumerate}
\item  $\omega=\dot r_\Sigma\,e^{(\lambda-\nu)/2}$ evaluated  on $r=r_{\Sigma}$.

\item  $T^\mu_{r;\mu}=0$ evaluated on $r=r_{\Sigma}$.

\item The equation relating the total mass loss rate with the energy flux
through the boundary surface.
\end{enumerate}
\item Depending on the kind of matter under consideration, the system of
surface equations described above may be closed with the additional
information provided by the transport equation
and/or the equation of state for the anisotropic pressure and/or
additional information about some of the physical variables evaluated on
the boundary surface (e.g. the
luminosity).

\item Once the system of surface equations is closed, it may be integrated for
any particular initial data.

\item  Feeding back the result of integration in the expressions for $m$ and
$\nu$, these two functions are completely determined.

\item  With the input from the point 7 above, and using field equations,
together with the equations of state and/or transport equation, all
physical variables may be found for any piece of
matter distribution.
\end{enumerate}

If the system is ``very far'' from equilibrium then it  could  be necessary to go  through the process once again by replacing the seed solution by the solution obtained  from the point 8 above. This could be done as many  times as it is required to obtain a satisfactory description of the system under consideration.
\section{WHAT HAPPENS IF WE RELAX THE PASCAL PRINCIPLE? (AND WHY?)}
Local isotropy of pressure (Pascal principle) is a common assumption in the study of self--gravitating systems, whenever the fluid approximation is used to describe the matter distribution of the object. This Pascalian character of fluids is supported by large amount of observational evidence pointing  towards the equality of  principal stresses under variety of circumstances. However strong  theoretical evidence presented in the last decades by  different authors suggest that  for certain density ranges, different kind of physical phenomena may take place, giving rise to local anisotropy (see \cite{4}--\cite{14N} for details and references). Among them we may mention:

\begin{enumerate}
\item ``Exotic'' phase transition (e. g. pion condensate).
\item\ Magnetic fields.
\item  Type II superconductor.
\item Type P superfluid.
\item Boson stars.
\item Viscosity.
\item Anisotropic velocity distributions.
\item Two fluid systems.
\item Gravstars.
\end{enumerate}

Besides the physical causes that may produce local anisotropy, it is worth noticing that General Relativity in some sense  introduces  an ``intrinsic'' anisotropy of pressure, by  assigning different roles to   radial ($P_r$) and tangential pressure ($P_\bot$). Indeed, let us take a look at eq.(14). Its  physical interpretation  is quite simple. The  first term on the left ( the gradient of pressure) equals the gravitational ``force'' term plus a term produced by the anisotropy. Now, the amazing point is that the pressure which appears in the gravitational term is the radial pressure. In other words, tangential pressure does not produce any regenerative pressure effect. Thus, in principle, one could obtain more massive configurations in equilibrium by increasing the tangential pressure, since the latter does not affects   the gravitational term.

Finally, the following question appears naturally:
How do the properties of the locally anisotropic system differ from the locally isotropic one?

As it could be expected they differ in many ways. However there are two which deserve particular attention:

\begin{enumerate}
\item Cracking induced by perturbations of local isotropy of pressure.  The concept of cracking was introduced in  \cite{c1} to describe the behaviour of spherically symmetric fluid distributions just after  their departure from equilibrium, when total nonvanishing radial forces of different signs appear within the system. We say that there is  cracking whenever this radial force is directed inward in the inner part of the sphere and reverses its sign beyond  some value of the radial coordinates (interior to the boundary of the fluid distribution). For the configurations considered in \cite{c1}, it was shown that cracking will occur only if deviations from local isotropy are allowed.  Furthermore in  \cite{c2} it was shown that for a large class of fluid distributions, cracking would appear if only small relative anisotropy is permitted ($\frac{\left(P_\bot-P_r\right)}{P_r}<<1$). 

It goes without saying that the appearance of cracking could drastically affect the evolution of the configuration.

\item Changes in the total mass allowed for a compact object. From the comments above about the role of  radial and tangential pressure, and the absence of a regenerative effect for the latter, it follows that the total mass allowed for a compact object would be seriously affected by the presence of local anisotropy.

\end{enumerate}

\section{HOW DO ELECTRIC CHARGE AND DISSIPATION AFFECT THE EVOLUTION OF MASSIVE STARS?}

We shall now illustrate the effects produced by dissipation and electric charge  in the process of gravitational collapse.  For this purpose we shall use the formalism  developed by Misner and collaborators based in comoving coordinates \cite{M}, \cite{Misner} (observe the change in notation and signature with respect to \cite{B})).

Thus, let us consider a spherically symmetric distribution  of collapsing
fluid, bounded by a spherical surface $\Sigma$. The fluid is
assumed to be locally anisotropic  and undergoing dissipation in the
form of heat flow (to model dissipation in the diffusion approximation), null radiation (to model dissipation in the free streaming approximation) and shearing
viscosity.

Choosing comoving coordinates inside $\Sigma$, the general
interior metric can be written
\begin{equation}
ds^2=-A^2dt^2+B^2dr^2+R^2(d\theta^2+\sin^2\theta d\phi^2),
\label{1}
\end{equation}
where $A$, $B$ and $R$ are functions of $t$ and $r$ and are assumed
positive. We number the coordinates $x^0=t$, $x^1=r$, $x^2=\theta$
and $x^3=\phi$.

The matter energy-momentum $T_{\alpha\beta}$ inside $\Sigma$
has the form
\begin{equation}
T_{\alpha\beta}=(\mu +
P_{\perp})V_{\alpha}V_{\beta}+P_{\perp}g_{\alpha\beta}+(P_r-P_{\perp})\chi_{
\alpha}\chi_{\beta}+q_{\alpha}V_{\beta}+V_{\alpha}q_{\beta}+
\epsilon l_{\alpha}l_{\beta}-2\eta\sigma_{\alpha\beta}, \label{3}
\end{equation}
where $\mu$ is the energy density, $P_r$ the radial pressure,
$P_{\perp}$ the tangential pressure, $q^{\alpha}$ the heat flux,
$\epsilon$ the energy density of the null fluid describing dissipation in the free streaming approximation, $\eta$ the
shear viscosity coefficient, $V^{\alpha}$ the four velocity of the fluid,
$\chi^{\alpha}$ a unit four vector along the radial direction
and $l^{\alpha}$ a radial null four vector. These quantities
satisfy
\begin{equation}
V^{\alpha}V_{\alpha}=-1, \;\; V^{\alpha}q_{\alpha}=0, \;\; \chi^{\alpha}\chi_{\alpha}=1, \;\;
\chi^{\alpha}V_{\alpha}=0, \;\; l^{\alpha}V_{\alpha}=-1, \;\; l^{\alpha}l_{\alpha}=0.
\end{equation}

Observe that we have assumed the shear viscosity  tensor $\pi_{\alpha \beta}$ to satisfy the relation \begin{equation}
\pi_{\alpha \beta}=-2\eta \sigma_{\alpha \beta},
\label{sv}
\end{equation}
where   $\sigma_{\alpha \beta}$ is the shear tensor.  However this last equation is valid only within the context of the standard irreversible thermodynamics (see \cite{8N}, \cite{FC} for details).

In a full causal picture of dissipative variables we should not assume (\ref{sv}). Instead, we should  use the  transport equation derived from the corresponding theory (e.g. the M\"{u}ller--Israel--Stewart theory \cite{Muller67}--\cite{IsSt76}). However for the sake of simplicity, in this work we shall restrict ourselves to the  standard irreversible thermodynamics  theory (only in what concerns shear viscosity, for the heat flux we shall consider a causal theory).

The acceleration $a_{\alpha}$ and the expansion $\Theta$ of the fluid are
given by
\begin{equation}
a_{\alpha}=V_{\alpha ;\beta}V^{\beta}, \;\;
\Theta={V^{\alpha}}_{;\alpha}. \label{4b}
\end{equation}
and its  shear $\sigma_{\alpha\beta}$ by
\begin{equation}
\sigma_{\alpha\beta}=V_{(\alpha
;\beta)}+a_{(\alpha}V_{\beta)}-\frac{1}{3}\Theta h_{\alpha \beta},\label{4a}
\end{equation}
where $h_{\alpha \beta}=g_{\alpha\beta}+V_{\alpha}V
_{\beta}
.$

Since we are interested in the influence of electric charge too, we need to include the electromagnetic energy tensor and the Maxwell equations.

The former is given by 
\begin{equation}
E_{\alpha\beta}=\frac{1}{4\pi}\left({F_{\alpha}}^{\gamma}F_{\beta\gamma}
-\frac{1}{4}F^{\gamma\delta}F_{\gamma\delta}g_{\alpha\beta}\right),
\label{6}
\end{equation}
where $F_{\alpha}^{\gamma}$ is the Maxwell tensor, which can be expressed through  the four vector potential as
\begin{equation}
F_{\alpha\beta}=\phi_{\beta,\alpha}-\phi_{\alpha,\beta}. \label{7}
\end{equation}
Since we are considering the charge to be comoving with the fluid, then there is no magnetic field, and the four density current is proportional to the four velocity
\begin{equation}
\phi_{\alpha}=\Phi\delta_{\alpha}^0, \;\; J^{\alpha}=\varsigma
V^{\alpha}, \label{9}
\end{equation}
where $\varsigma$ denotes the charge density. Also it  can be easily obtained that the total charge interior to a sphere labelled by coordinate $r$ is independent on time and given by,
\begin{equation}
s(r)=4\pi\int^r_0\varsigma BR^2dr. \label{13}
\end{equation}

From all the above we  may  write the Einstein--Maxwell equations

\begin{equation}
G_{\alpha\beta}=8\pi(T_{\alpha\beta}+E_{\alpha\beta}),
\label{2}
\end{equation}
\begin{eqnarray}
{F^{\alpha\beta}}_{;\beta}=4\pi J^{\alpha}. \label{8}
\end{eqnarray}

However we shall not need  those equations explicitly, and therefore we shall not present them here. Instead we shall use Bianchi identities.

\subsection{Dynamical equation}

There are two non
 trivial  components of the Bianchi identities, however 
for our purpose we only need one of them, which reads
 \begin{eqnarray}
&&(T^{\alpha\beta}+E^{\alpha\beta})_{;\beta}\chi_{\alpha}
=\frac{1}{A}(\dot{q}+\dot{\epsilon})
+\frac{1}{B}\left(P_r+\epsilon-\frac{4}{3}\eta\sigma\right)^{\prime}
 \nonumber\\
&&+2(q+\epsilon)\frac{\dot B}{AB}+2(q +\epsilon)\frac{\dot C}{AC} \nonumber\\
&&
 +\left(\mu+P_r+2\epsilon-\frac{4}{3}\eta\sigma\right)\frac{A^{\prime}}{AB}
\label{III18}\nonumber\\
&&+2(P_r-P_{\perp}+\epsilon-2\eta\sigma)\frac{(Cr)^{\prime}}{BCr}
 -\frac{ss^{\prime}}{4\pi B(Cr)^4}=0,
\end{eqnarray}
where $C\equiv R/r$.

It would  be convenient to introduce the proper time derivative
\begin{equation}
D_T=\frac{1}{A}\frac{\partial}{\partial t}, \label{16}
\end{equation}
and the $R$ derivative
\begin{equation}
D_R=\frac{1}{R^{\prime}}\frac{\partial}{\partial r}. \label{23a}
\end{equation}
Using (\ref{16}) we can define the velocity $U$ of the collapsing
fluid (for another definition of velocity see \cite{W} ) as the variation of the areal radius ($R$)  with respect to proper time, i.e.\

\begin{equation}
U=D_TR<0 \;\; \mbox{(in the case of collapse)}. \label{19}
\end{equation}
The mass function $m$ defined by 
\begin{equation}
m=\frac{R^3}{2}{R_{23}}^{23}+\frac{s^2}{2R}
=\frac{R}{2}\left\{\left(\frac{\dot{R}}{A}\right)^2
-\left(\frac{R^{\prime}}{B}\right)^2+1\right\}+\frac{s^2}{2R},
 \label{18}
\end{equation}
may also be written as 
\begin{equation}
m=\int^{R}_{0}4\pi R^2 \left[\mu +
\epsilon+(q+\epsilon)\frac{U}{E}\right]dR+\frac{s^2}{2R}+\frac{1}{2}\int^{R
}_{0}\frac{s^2}{R^2}dR
\label{27int}
\end{equation}
(assuming a regular centre to the distribution, so $m(0)=0$). Where 
\begin{equation}
E \equiv \frac{(R)^{\prime}}{B}=\left[1+U^2-\frac{2m(t,r)}{R}
+\left(\frac{s}{R}\right)^2\right]^{1/2}.
\label{20}
\end{equation}

Using all the relations above, we obtain after some lengthy calculations (see \cite{13} for details)
\begin{eqnarray}
\left(\mu+P_r+2\epsilon-\frac{4}{3}\eta\sigma\right) D_TU = \nonumber\\
-\left(\mu+P_r+2\epsilon-\frac{4}{3}\eta\sigma\right)
\left[m
+4\pi\left(P_r+\epsilon-\frac{4}{3}\eta\sigma\right)R^3
-\frac{s^2}{R}\right]\frac{1}{R^2}\nonumber \\
-E^2\left [D_R\left(P_r+\epsilon-\frac{4}{3}\eta\sigma\right)\right.
+2(P_r-P_{\perp}+\epsilon-2\eta\sigma)\frac{1}{R}\nonumber \\
\left.-\frac{s}{4\pi R^4}D_Rs\right]%\nonumber \\
-E\left[D_Tq+D_T\epsilon+4(q+\epsilon)\frac{U}{R}
 +2(q+\epsilon)\sigma\right].\nonumber \\
\label{eqd}
\end{eqnarray}
Equation  (\ref{eqd}) has the  ``Newtonian'' form 
\begin{equation}
Force= Mass \; (density) \times (radial) Acceleration
\label{Newton}
\end{equation}
which in the non-dissipative locally isotropic case coincides
with Eq.\ (43) in \cite{Bek}. Let us now analyze in some detail
the three terms on the right of (\ref{eqd}).

The first term  on the right hand side of (\ref{eqd}) represents the
gravitational force. The factor within the round bracket (the same factor
as on the left of (\ref{eqd})) defines the
inertial mass density (``passive''  gravitational mass density) and shows
how it is affected by dissipative terms. Observe that it is not affected by
the electric charge.

The factor within
the first square bracket shows how dissipation and the electric charge
affect the ``active'' gravitational mass term.  Using (\ref{27int}) in
(\ref{eqd}) we see that the charge
will increase the ``active gravitational mass'' only if
\begin{equation}
\int^{R}_{0}\frac{s^2}{R^2}dR>\frac{s^2}{R}
\label{con}
\end{equation}
otherwise it will decrease it. In the case  \begin{equation}
\int^{R}_{0}\frac{s^2}{R^2}dR=\frac{s^2}{R}  \label{conIII}
\end{equation}
no regenerative effect of charge occurs. This strange effect was already noticed by
Bekenstein \cite{Bek}, and enhances the possibility that  Coulomb repulsion
might prevent the gravitational
collapse of the sphere.

There are three different contributions in the second square bracket. The
first one is just the gradient of the total ``effective'' radial  pressure
(which includes the radiation
pressure and the influence of shear viscosity on $P_r$). The second
contribution comes from the local anisotropy of pressure, including the
contributions from the radiation pressure
and shear viscosity. Finally the last term describes Coulomb repulsion,
which is always positive (always opposing gravitation). Observe that this  repulsive effect is reinforced if the charged distribution is such that (\ref{con}) is violated, due to the ensuing decreasing of the ``active gravitational mass'' term.

The last square bracket contains different contributions due to dissipative
processes. The third term within this bracket is positive ($U<0$) showing
that the outflow of
$q>0$ and $\epsilon>0$ diminish the total energy inside the collapsing
sphere, thereby reducing the rate of collapse. The last term describes an
effect resulting from the coupling of
the dissipative flux with the shear of the fluid. The effects of
$D_T\epsilon$ have
been discussed in detail in
\cite{MisnerII}. Thus it  remains to analyse the effects of $D_Tq$;
this depends on the transport equation adopted, and we will next proceed to
study one case.

\subsection{Coupling the transport equation with the dynamical equation}
 We shall use a transport equation derived from the
M\"{u}ller-Israel-Stewart second
order phenomenological theory for dissipative fluids \cite{Muller67, IsSt76}.

Indeed, it is well known that the Maxwell-Fourier law for
radiation flux leads to a parabolic equation (diffusion equation)
which predicts propagation of perturbations with infinite speed
(see \cite{6}-\cite{8'} and references therein). This simple fact
is at the origin of the pathologies \cite{9} found in the
approaches of Eckart \cite{10bis} and Landau \cite{11bis} for
relativistic dissipative processes. To overcome such difficulties,
various relativistic
theories with non-vanishing relaxation times have been proposed in
the past \cite{Muller67,IsSt76,14,15}. The important point is that
all these theories provide a heat transport equation which is not
of Maxwell-Fourier type but of Cattaneo type \cite{18}, leading
thereby to a hyperbolic equation for the propagation of thermal
perturbations.

Thus we shall  assume that our heat flux vector satisfies:
\begin{equation}
\tau
h^{\alpha\beta}V^{\gamma}q_{\beta;\gamma}+q^{\alpha}=-\kappa h^{\alpha\beta}
(T_{,\beta}+Ta_{\beta}) -\frac 12\kappa T^2\left( \frac{\tau
V^\beta }{\kappa T^2}\right) _{;\beta }q^\alpha ,  \label{21t}
\end{equation}
where $\tau$ denotes the relaxation time and $\kappa$ is the coefficient of thermal conductivity.

Then, coupling (\ref{21t}) with (\ref{eqd}) one obtains after some calculations (see \cite{13} for details)
\begin{eqnarray}
&&\left(\mu+P_r+2\epsilon-\frac{4}{3}\sigma\eta\right)(1-\alpha)D_TU
 \nonumber \\
&=&(1-\alpha)F_{grav}+F_{hyd}+\frac{\kappa E^2}{\tau}D_RT \nonumber \\
&+&E\left[\frac{\kappa T^2q}{2\tau}D_T\left(\frac{\tau}{\kappa
   T^2}\right)-D_T\epsilon\right]
\nonumber \\
&-&Eq\left(\frac{5}{2}\frac{U}{R}+\frac{3}{2}\sigma-\frac{1}{\tau}\right)
 -2E\epsilon\left(2\frac{U}{R}+\sigma\right),
\label{V4}
\end{eqnarray}
with 
\begin{eqnarray}
F_{grav}&&=-\left(\mu+P_r+2\epsilon -\frac{4}{3}\eta
\sigma\right)\nonumber\\
&\times&
\left[m+4\pi\left(P_r+\epsilon-\frac{4}{3}\eta \sigma
\right)R^3-\frac{s^2}{R}\right]\frac{1}{R^2},
\label{grav}\\
F_{hyd}&=& -E^2 \left[D_R \left(P_r+\epsilon -\frac{4}{3}\eta
\sigma\right) \right. \nonumber\\
&+&\left.2(P_r-P_{\perp}+\epsilon-2\eta\sigma)\frac{1}{R}-\frac{s}{4\pi
R^4}D_Rs\right], \label{hyd}
\end{eqnarray}
and 
\begin{equation}
\alpha=\frac{\kappa
T}{\tau}\left(\mu+P_r+2\epsilon-\frac{4}{3}\sigma\eta\right)^{-1}.
\label{alpha}
\end{equation}
The first observation to make is that once the transport equation has been taken into account, then the
inertial energy density  and the ``passive gravitational mass density'',
i.e the factor multiplying $D_TU$ and
the first factor at the right of (\ref{grav}) respectively (which of course
are the same, as expected from the equivalence principle), appear
diminished by the factor $1-\alpha$, a result
already obtained, but here generalized by the inclusion of
the viscosity and radiative phenomena (see \cite{8u}--\cite{10} and references therein). Also observe that the charge does not enter into the definition of $\alpha$.
However it does affect the ``active gravitational mass'' (the factor within
the square bracket in
(\ref{grav})).

Let us now consider a system which starts to collapse.  As far as the right hand side of (\ref{V4}) is negative, the
system
keeps collapsing. However, let us assume that the collapsing sphere evolves  in such a way that, for some region of the sphere, the value of
$\alpha$ increases  and  approaches the
critical value of
$1$. Then, as this process goes on, the ensuing decreasing of the
gravitational force term would eventually lead to a change of the sign
of the right hand side of (\ref{V4}). Since that would happen for small
values of the
effective inertial mass density, that would imply a strong bouncing of that
part of the  sphere, even for a small absolute value of the
right hand side of (\ref{V4}).
Obviously for this scenario  to occur, evolution should proceed in such a way that $\alpha$ approaches the critical value of $1$.
At present we may speculate that  $\alpha$ may
increase substantially (for non-negligible values of $\tau$) in a pre-supernovae event.

Indeed, at the last stages of massive star evolution, the decreasing of the opacity
of the fluid, from very high values preventing the propagation of photons
and neutrinos   \cite{Ar}, to smaller values, gives rise to neutrino
radiative heat conduction. Under these conditions both $\kappa$ and $T$
could be sufficiently large as to imply a substantial increase of
$\alpha$. In fact, the values suggested in \cite{Ma} ($[\kappa] \approx
10^{37}$;
$[T] \approx 10^{13}$; $[\tau] \approx 10^{-4}$; $[\rho] \approx
10^{12}$, in c.g.s. units and Kelvin) lead to $\alpha \approx 1$. 

 A numerical model illustrating this effect has  recently been presented.
\cite{11u}

Before closing this section it is worth mentioning that the reported effect (the decreasing of the ``passive gravitational mass density term'' by a   factor ($1-\alpha$), is related to  the first term on the left and the $Ta_\mu$ term, in (\ref{21t}). But these terms have to be present in any causal theory of dissipation. Accordingly such an effect is also
expected to hold for a general family of theories which includes the  M\"{u}ller-Israel-Stewart theory. 
\section{Conclusions}
To summarize:

\begin{itemize}
\item  Dissipative phenomena play a relevant role in the dynamics of collapse. In particular, relaxational effects may drastically change the outcome of gravitational collapse.
\item The dynamical regime may be approached by means of successive approximations. Doing so, we are able to describe the evolution by  solving ordinary differential equations, instead of partial differential equations.
\item Local anisotropy of pressure has to be taken into  consideration in the study of the structure and  evolution  of  massive stars.
\item Not only the absolute value of electric charge but also its distribution may be relevant in stellar structure and evolution.
\end{itemize}


\begin{thebibliography}{88}
\bibitem{HJR} L. Herrera, J. Jim\'enez and  G. Ruggeri {\it Phys.Rev.D} {\bf 22}, 2305(1980).
\bibitem{FCP} L. Herrera and L. N\'u\~nez {\it Fundamental of Cosmic Physics}, {\bf 14} , 235 (1990).
\bibitem{5bis}L. Herrera, W. Barreto, A. Di Prisco and N.O. Santos
{\it Phys. Rev. D} {\bf 65}, 104004 (2002).
\bibitem{B} H. Bondi {\it Proc. R. Soc. London} {\bf A281}, 39 (1964).
\bibitem{4} L. Herrera and N. O. Santos  {\it Phys. Rep.} {\bf 286}, 53 (1997).  
\bibitem{9} L. Herrera, A.  Di Prisco, J. Mart\'\i n, J. Ospino, N. O.
Santos and O. Troconis  {\it Phys. Rev. D} {\bf 69}, 084026 (2004).
\bibitem{14N} L. Herrera, J. Ospino and  A. Di Prisco 
{\it  Phys. Rev. D} {\bf 77}, 027502, (2008).

\bibitem {c1}  L. Herrera {\it Phys. Lett. A} {\bf 165}, 206, (1992).

\bibitem{c2} A. Di Prisco, L. Herrera and  V. Varela.
{\it Gen. Rel. Grav.}  {\bf 29}, 1239, (1997).
\bibitem{M} C.W. Misner and D. H. Sharp {\it Phys. Rev.} {\bf 136}, B571 (1964).
\bibitem{Misner} C.W. Misner {\it Phys. Rev.} {\bf 137}, B1360  (1965).
\bibitem{8N} R.  Maartens, {\it astro-ph}/9609119.
\bibitem{FC} L. Herrera, A. Di Prisco, E. Fuenmayor and   O. Troconis,
{\it Int. J.  Mod. Phys.  D} {\bf 18}, 129, (2009).
\bibitem{Muller67}  I M\"{u}ller {\it Z. Physik} {\bf 198}, 329 (1967)
\bibitem{IsSt76} W  Israel {\it Ann. Phys.} (NY) {\bf 100}, 310 (1976); W
Israel and J Stewart {\it Phys. Lett. A} {\bf 58},
213  (1976); {\em Ann. Phys.} (NY) {\bf 118}, 341 (1979)
\bibitem{W} L. Herrera, N. O. Santos and A. Wang {\it Phys.Rev. D} {\bf 78}, 084026 (2008).
\bibitem{13} A Di Prisco, L. Herrera, G. Le Denmat, M. MacCallum and   N.O. Santos {\it Phys. Rev. D} {\bf 76}, 064017, (2007). 
\bibitem{Bek} J Bekenstein  {\it Phys. Rev. D} {\bf 4}, 2185 (1971)
\bibitem{MisnerII} C Misner {\it Phys. Rev.} {\bf 137}, B1360 (1965)


\bibitem{6} D Joseph  and L Preziosi {\it Rev. Mod. Phys.}
{\bf 61}, 41 (1989)

\bibitem{7} J Casas-V\'azquez and G Lebon {\it Rep. Prog. Phys.}
{\bf 51}, 1105 (1988)

\bibitem{8'} L Herrera and D Pav\'on
{\it Physica A}, {\bf 307}, 121 (2002)

\bibitem{9} W Hiscock  and L Lindblom {\it Ann. Phys.} (NY)
{\bf 151}, 466 (1983)

\bibitem{10bis} C Eckart {\it Phys. Rev.} {\bf 58}, 919 (1940)

\bibitem{11bis} L Landau and E Lifshitz, {\it Fluid Mechanics}
(Pergamon Press, London) (1959)

\bibitem{14} D Pav\'on, D Jou  and J Casas-V\'azquez {\it Ann. Inst. H
Poincar\'e} {\bf A36}, 79 (1982)

\bibitem{15} B Carter {\it Journ\'ees Relativistes}, ed. M Cahen, R
Debever and J Geheniau, (Universit\'e Libre de Bruxelles) (1976)

\bibitem{18} C Cattaneo {\it Atti Semin. Mat. Fis. Univ. Modena}
{\bf 3}, 3 (1948)
\bibitem{8u} L. Herrera
{\it Phys. Lett. A}, {\bf 300}, 157  (2002).

\bibitem{10} L. Herrera {\it Int. J. Mod. Phys. D} {\bf 15}, 2197 (2006).


\bibitem{Ar} W. Arnett W  {\it Astrophys. J.}, {\bf 218}, 815 (1977).

\bibitem{Ma}J.  Mart\'\i nez {\it Phys. Rev. D}, {\bf 53}, 6921 (1996).
\bibitem{11u} L. Herrera, A. Di Prisco and  W. Barreto {\it Phys, Rev. D} {\bf 73}, 024008 (2006). 

\end{thebibliography}
\end{document}